\documentclass[useAMS,usenatbib,usegraphicx]{mn2e}
\onecolumn{}

\title[WASP 1814+48]{The Pre-He White Dwarfs in Eclipsing Binaries. IV. WASP 1814+48 with Multiperiodic Pulsations}

\author[J. W. Lee et al.]
       {Jae Woo Lee$^{1}$\thanks{E-mail: jwlee@kasi.re.kr}, Kyeongsoo Hong$^{2}$, Hye-Young Kim$^{3}$ and Jang-Ho Park$^{1}$ \\
        $^1$Korea Astronomy and Space Science Institute, Daejeon 34055, Republic of Korea \\
        $^2$Institute for Astrophysics, Chungbuk National University, Cheongju 28644, Republic of Korea \\ 
        $^3$Department of Astronomy and Space Science, Chungbuk National University, Cheongju 28644, Republic of Korea}

\begin{document}

\date{Accepted 2022 ---------. Received 2022 ---------; in original form 2022 }

\pagerange{\pageref{firstpage}--\pageref{lastpage}} \pubyear{2021}

\maketitle

\label{firstpage}

\begin{abstract}
For the EL CVn candidate 1SWASPJ181417.43+481117.0 (WASP 1814+48), we secured the first spectroscopic observations between 
2015 April and 2021 March. Using the echelle spectra, the radial velocities (RVs) of the primary star were measured with 
its atmospheric parameters of $T_{\rm eff,1}=7770\pm130$ K and $v_1$$\sin$$i=47\pm6$ km s$^{-1}$. We fitted our single-lined RVs 
and the TESS light curve simultaneously. From the binary modeling, we determined the following fundamental parameters for 
each component: $M_1=1.659\pm0.048$ $M_\odot$, $R_1=1.945\pm0.027$ $R_\odot$, and $L_1=12.35\pm0.90$ $L_\odot$ for WASP 1814+48 A, 
and $M_2=0.172\pm0.005$ $M_\odot$, $R_2=0.194\pm0.005$ $R_\odot$, and $L_2=0.69\pm0.07$ $L_\odot$ for WASP 1814+48 B. The surface 
gravity of $\log g_2=5.098\pm0.026$ obtained from $M_2$ and $R_2$ is concurrent with 5.097$\pm$0.025 computed directly from 
the observable quantities. WASP 1814+48 B is well-matched with the 0.176 M$_\odot$ white dwarf (WD) evolutionary model 
for $Z=0.01$. The metallicity and our Galactic kinematics indicate that the program target is a thin-disk star. The whole light 
residuals after removal of the binary trend were analyzed and found to oscillate at a total of 52 frequencies. Among these, 
most of the low frequencies below 24 day$^{-1}$ are aliases and orbital harmonics. The five significant frequencies between 32 
and 36 day$^{-1}$ are the pulsation modes of WASP 1814+48 A located in the $\delta$ Sct domain on ZAMS, and the high frequencies 
of 128$-$288 day$^{-1}$ arise from WASP 1814+48 B in the pre-He WD instability strip. Our results reveal that WASP 1814+48 is 
the fifth EL CVn star that is composed of a $\delta$ Sct-type primary and a pre-ELMV (extremely low-mass pre-He WD variable). 
\end{abstract}


\section{INTRODUCTION}

Extremely low-mass ($\la 0.3$ M$_\odot$) white dwarfs (ELM WDs) are the remnants of stars that cannot burn helium (He) in 
their cores. Because the Universe does not have enough time to produce them by single-star evolution, the He-core ELM WDs 
are thought to be formed by considerable mass loss at the red giant branch phase of stellar binaries before He burning 
(Marsh, Dhillon \& Duck 1995; Kilic, Stanek \& Pinsonneault 2007). In this case, they can provide information about 
the past evolution of the precursor binaries. We have been performing time-series spectroscopy on eclipsing binaries (EBs) 
containing an ELM WD precursor (pre-ELM WD) as possible (Lee et al. 2020). The main purpose of the observations is to measure 
the fundamental parameters of the interesting rare objects, combining existing or new photometric data, and to present 
their evolution scenario. The promising targets for this subject are EL CVn-type binaries, which are post-mass transfer stars 
comprising an A/F main sequence and a pre-ELM WD companion (Maxted et al. 2014; van Roestel et al. 2018). 

Five out of $\sim70$ EL CVn stars are pulsating EBs that exhibit possible multiperiodic oscillations arising from both 
a $\delta$ Sct (or $\gamma$ Dor)-type primary and a pre-ELMV companion (Hong et al. 2021; Kim et al. 2021). Because 
the fundamental parameters such as masses and radii can be measured accurately and in detail, spectroscopic and 
eclipsing binaries provide us with an opportunity to test and refine the evolutionary models of stars (Hilditch 2001; 
Torres, Andersen \& Gim\'enez 2010). At the same time, pulsating stars help to probe and constrain their interior physics 
from core to surface through asteroseismology (Antoci et al. 2019; Aerts 2021). Thus, the stellar pulsations in EBs are 
of great value, because the synergy between the two kinds of variables can clearly enhance our understanding of 
stellar physics (Murphy 2018; Kurtz 2022).

We focus on the EL CVn-type star 1SWASPJ181417.43+481117.0 (WASP 1814+48; TIC 420947520; Gaia EDR3 2122136709525411072), 
which was recognized by Maxted et al. (2014) to be an eclipsing variable with a period of 1.7994305 $\pm$ 0.0000005 days. 
From archival WASP photometry, they found that the binary target is a totally-eclipsing detached system with a mass ratio 
of $q$ = 0.134 $\pm$ 0.015, relative radii of $r_1$ = 0.2565 $\pm$ 0.0025 and $r_2$ = 0.0247 $\pm$ 0.0004, and 
surface brightness and luminosity ratios of $J$ = 3.000 $\pm$ 0.064 and $L_2/L_1$ = 0.0277 $\pm$ 0.0002, respectively. 
Further, the effective temperatures of the brighter, more massive primary star and its companion were obtained to be 
$8000\pm300$ K and $12,500\pm1800$ K, respectively, by comparing the observed and synthetic flux distributions. This work 
is the fourth in a paper-series on the pre-ELM WDs in EBs (Lee et al. 2020; Lee, Hong \& Park 2022; Hong et al. 2021). 
We analyze in detail our high-resolution spectra and the TESS photometric data of WASP 1814+48, and report the discovery 
of multiple types of pulsations originating from the EB system.

\section{TESS PHOTOMETRY AND ECLIPSE TIMINGS}

Highly precise photometry of WASP 1814+48 has been performed by the TESS mission (Ricker et al. 2015) from 2019 July 18. 
We downloaded the 2-min cadence data taken during Sector 14, 25-26, and 40-41 from MAST\footnote{https://archive.stsci.edu/} 
and used the simple aperture photometry (\texttt{SAP$_-$FLUX}) data in this study. The flux measurements were detrended and 
normalized by fitting a second-order polynomial to the out-of-eclipse portion of each sector's light curve, and they were 
converted to magnitude units. The resultant observations are displayed in the top panel of Figure 1. A total of 91,240 
individual points were obtained for the five sectors. The crowdedness factor CROWDSAP reported in the TESS data is the ratio 
of the target flux to the total flux in a photometric aperture. This can be used to see if nearby stars are observed in 
the same pixel as the target, where a value of 1 means that there is no contamination. The CROWDSAP value for WASP 1814+48 
is 0.9900$\pm$0.0024 on average for these sectors.

The TESS eclipse times of WASP 1814+48 and their errors were determined using the Kwee \& van Woerden (1956) method. These 
are presented in Table 1, where Min I and II present the primary and secondary eclipses, respectively, at orbital phases 0.0 
and 0.50. In order to obtain the updated linear ephemeris of the binary star and to phase its time-series data, we used 
the primary minimum epochs in the following least-squares solution:
\begin{equation}
\mbox{Min I} = \mbox{BJD}~ 2,459,010.506694(27) + 1.79943078(17)E.
\end{equation}
The 1$\sigma$-error values for each coefficient are given in the parentheses. The timing residuals from the linear ephemeris 
appear as $O-C$ in Table 1. The TESS light curve phased with Equation (1) is depicted in the middle panel of Figure 1.

\section{GROUND-BASED SPECTROSCOPY AND DATA ANALYSIS}

The spectroscopic observations of WASP 1814+48 were performed with the Bohyunsan Observatory Echelle Spectrograph 
(BOES; Kim et al. 2007), attached to the 1.8-m telescope in the Bohyunsan Optical Astronomy Observatory (BOAO), Korea. 
A total of 31 spectra were secured for seven nights between 2015 April and 2021 March. Their wavelength coverage ranges from 
3600 to 10,200 $\rm \AA$ with a resolving power of $R=30,000$. Each exposure time of our target star was 40 min, resulting 
in a signal-to-noise (S/N) ratio of approximately 20 around 4500 $\rm \AA$. This corresponds to 0.015 of the eclipsing period, 
so orbital smearing was not considered. All observed spectra were reduced by following the IRAF standard data-reduction 
procedures (Hong et al. 2015), which include flat fielding, de-biasing, extraction, and wavelength and flux calibration.

Figure 2 shows the trailed spectra of WASP 1814+48 in the Mg II $\lambda$4481 region. In this figure, the S-wave feature 
of the more massive primary component (WASP 1814+48 A) is clearly shown, while there is no sign of the hotter secondary star 
(WASP 1814+48 B). To measure the radial velocities (RVs) from the observed spectra, we used the cross-correlation function 
(CCF) method (Simkin 1974; Tonry \& Davis1979) implemented in the RaveSpan software (Pilecki, Konorski \& Gorski 2012; 
Pilecki et al. 2017). In the run, most of metallic and Balmer lines are difficult to measure, so we selected the spectral region 
of Mg II $\lambda$4481 that is useful for determining the RVs of the EL CVn-type stars (Lee et al. 2020 for WASP 0131+28; 
Hong et al. 2021 for WASP 0843-11; Lee, Hong \& Park 2022 for WASP 1625-04). Template spectra were taken from 
the BOSZ stellar atmosphere models\footnote{https://archive.stsci.edu/prepds/bosz} (Bohlin et al. 2017) matching the effective 
temperature ($T_{\rm eff,1}$) and surface gravity ($\log g_1$) of the primary component discussed below. We measured the RVs of 
WASP 1814+48 A and they are presented in Figure 3 and Table 2. The RV measurements were fitted to a sine wave (Lee et al. 2018), 
in order to obtain the spectroscopic orbit of WASP 1814+48 A. The results from this calculation are summarized in Table 3, 
where $\gamma$ is the systemic velocity, $K_1$ and $a_1$$\sin$$i$ are the velocity semi-amplitude and semimajor axis of 
the primary component, and $f$(M) is the mass function. The orbital period $P$ = 1.79943078 days was taken from Eq. (1). 

To determine the rotational velocities ($v_1$$\sin$$i$) and surface temperature ($T_{\rm eff,1}$) of WASP 1814+48 A, we selected 
five absorption lines, Ca II K $\lambda$3933, H$_{\rm \gamma}$ $\lambda$4340, Fe II $\lambda$4383, H$_{\rm \beta}$ $\lambda$4861, 
and H$_{\rm \alpha}$ $\lambda$6563, which are useful spectral indicator for A0$-$F0 type stars (Gray \& Corbally 2009). 
From the BOES spectra, each absorption region was combined using the FDB\textsc{inary} code (Iliji\'c et al. 2004) to obtain 
a better S/N spectrum. Maxted et al. (2014) estimated the surface temperature of WASP 1814+48 A to be $8000\pm300$ K, based on 
the surface brightness ratio and the observed flux distribution. Therefore, the reconstructed spectrum was compared to 
the synthetic spectra with a temperature range of 6500 K$-$9000 K (in steps of 10 K) and the projected rotational velocity range 
of 10$-$200 km s$^{-1}$ (in steps of 1 km s$^{-1}$). The synthetic spectra were interpolated using the stellar models from 
the BOSZ spectral library (Bohlin et al. 2017) by adopting the solar metallicity and the surface gravity of $\log g_1$ = 4.1 
(cf. Section 4). We obtained both atmospheric parameters of WASP 1814+48 A by performing a $\chi^2$ grid search that minimizes 
the difference between the synthetic and reconstructed spectra and averaging the values found in each region (Hong et al. 2017; 
Lee, Hong \& Park 2022). As a consequence, $T_{\rm eff,1}=7770 \pm 130$ K and $v_1$sin$i=47\pm6$ km s$^{-1}$. In Figure 4, 
the reconstructed spectrum from the FDB\textsc{inary} code is presented with the best-fit model.

\section{BINARY MODELING} 

As with the archival WASP observations, the TESS light curve of WASP 1814+48 shows a box-shaped primary eclipse and 
an ellipsoidal variation. The shape resembles that of EL CVn, its class prototype, and there are no noticeable differences 
between each sector. The observed depths for the primary and secondary eclipses are 0.031 mag and 0.016 mag, respectively, 
in the TESS photometry, and 0.037 mag and 0.017 mag in the WASP one. In both the WASP and TESS light curves, 
the phase difference between Min I and Min II implies that WASP 1814+48 is in a circular orbit. To obtain improved 
binary parameters, we solved the TESS photometric data with our RV curve using the detached mode 2 of the Wilson-Devinney 
program (Wilson \& Devinney 1971; van Hamme \& Wilson 2007; hereafter W-D) and applied a mass ratio ($q$)-search method 
(e.g., Lee, Hong \& Kristiansen 2019). The binary modeling was done in a manner similar to the single-lined eclipsing system 
HW Vir (Lee et al. 2009) and the EL CVn-type star WASP 0131+28 (Lee et al. 2020). 

In the synthesis of the RV and light curves, the surface temperature of WASP 1814+48 A was given as $T_1$ = 7700 $\pm$ 130 K 
from our spectroscopic analysis, as discussed in the previous section. The albedos ($A$) and gravity-darkening exponents ($g$) 
for both components were all set to 1.0 from their temperatures. The logarithmic limb-darkening (LD) law was adopted and 
its bolometric ($X$, $Y$) and monochromatic ($x$, $y$) coefficients were interpolated from the values of van Hamme (1993) 
incorporated into the W-D program. Because the TESS passband is not included in the binary code, we used the LD parameters for 
Cousins $I_{\rm c}$-band. Furthermore, a circular orbit ($e$ = 0) and a synchronous rotation ($F_1 = F_2$ = 1.0) were applied. 
In this article, WASP 1814+48 A and B are denoted by subscripts 1 and 2, respectively.

The mass ratio is considered the most important parameter in binary modeling. We can directly calculate the velocity 
semi-amplitudes ($K_1$ and $K_2$) and the mass ratio ($q=K_1/K_2$) from double-lined RV curves. However, WASP 1814+48 B was 
too faint to measure its RVs. Hence, we computed a series of binary models for assumed mass ratios below 0.3, the so-called 
$q$-search procedure. The sum of squares of residuals ($\sum W(O-C)^2$) showed a global minimum around $q$ = 0.104. 
The $q$ value was adjusted to obtain the best-fit parameters of the eclipsing system. The final result is presented in 
Table 4, and the synthetic light and RV curves are displayed as solid curves in the middle panel of Figure 1 and 
the upper panel of Figure 3, respectively. The residuals between the observations and the binary model are plotted in 
the bottom panels of each figure. We can see our modeling result that describes the light and RV curves very well. 
The parameters' errors in Table 4 were obtained following the method introduced by Southworth et al. (2020) and applied by 
Lee, Hong \& Kim (2021) for the highly-precise TESS data. 

The fundamental stellar parameters of WASP 1814+48 A and B were first computed from our simultaneous light and velocity solution. 
The results are summarized in Table 5. We used the solar temperature and bolometric magnitude of $T_{\rm eff}$$_\odot$ = 5780 K 
and $M_{\rm bol}$$_\odot$ = +4.73, respectively. The bolometric corrections (BCs) were made using the empirical calibration 
between $\log T_{\rm eff}$ and BC presented in Flower (1996) and later corrected in Torres (2010). 
Using an apparent visual magnitude of $V$ = 10.673 $\pm$ 0.008 and the color excess of $E$($B-V$) = 0.028 $\pm$ 0.007 
(Stassun et al. 2019), we determined the distance of the target star to be 536 $\pm$ 20 pc. Within 3$\sigma$-error values, 
our distance coincides with 586 $\pm$ 5 pc inverted from the GAIA EDR3 parallax of 1.706 $\pm$ 0.013 mas 
(Gaia Collaboration et al. 2021) and 584 $\pm$ 4 pc estimated from the GAIA EDR3 measurements by Bailer-Jones et al. (2021). 

The synchronous rotations for the component stars were calculated to be $v_{\rm 1,sync} = 54.70 \pm 0.77$ km s$^{-1}$ and 
$v_{\rm 2,sync} = 5.46 \pm 0.15$ km s$^{-1}$ from the binary period $P$ and their radii $R_{1,2}$. Since both values of 
$v_{\rm 1,sync}$ and $v_1$$\sin$$i$ coincide close to the margins of errors, we believe the primary star is currently in 
a state of synchronized rotation or nearly so.

\section{PULSATIONAL CHARACTERISTICS}

The physical parameters of WASP 1814+48 in Table 5 indicate that the primary and secondary components are $\delta$ Sct and 
pre-ELMV candidates, respectively (Maxted et al. 2013; Wang, Zhang \& Dai 2020; Hong et al. 2021). To find the pulsation signatures 
present in the TESS observations, we applied a multifrequency analysis to the entire light curve residuals from our binary model. 
The PERIOD04 software of Lenz \& Breger (2005) was conducted up to the Nyquist limit of about 360 day$^{-1}$. The periodogram 
for WASP 1814+48 is presented in Figure 5, where two dominant signals are clearly visible around 33 day$^{-1}$. In addition, 
as shown in the inset box, there are the orbital frequency ($f_{\rm orb}$ = 0.55573 day$^{-1}$) and its multiples ($Nf_{\rm orb}$, 
$N$ is an integer) in the low-frequency domain. We performed an iterative pre-whitening process (Lee et al. 2014), and extracted 
52 frequencies with an S/N amplitude ratio larger than about 4.0 (Breger et al. 1993). The results of this process are summarized 
in Table 6, and the uncertainties of each parameter were computed following the method proposed by Kallinger, Reegen \& Weiss (2008). 
The amplitude spectra after pre-whitening the first two frequencies and then 38 frequencies are illustrated in the middle and 
bottom panels of Figure 5, respectively.

Of the extracted signals, possible orbital harmonics and combination frequencies were carefully examined using the frequency 
resolution of $\Delta f \simeq$ 0.002 day$^{-1}$ (Loumos \& Deeming 1978). They are remarked in the last column of Table 6. 
Most frequencies lower than 24 day$^{-1}$ may be either alias effects or orbital harmonics up to $42f_{\rm orb}$. The integer 
multiples of the orbital frequency can arise from tidally excited pulsations by the companions in eccentric binary systems 
such as the heartbeat star KOI-54 (Welsh et al. 2011). However, our binary model favors the conclusion that WASP 1814+48 is in 
a circular orbit. Thus, it is difficult to regard the observed multiples of $Nf_{\rm orb}$ as stellar oscillations excited by 
tidal interaction. We think the aliases and the orbital harmonics result from insufficient removal of the binary effects and 
the systematic trends in the TESS data.

To cross-check the significant frequencies found from the entire TESS observations and to examine their variations with time, 
we independently analyzed the light curve residuals of each sector in the same way as before. The resultant frequencies were 
compared with those in Table 6. Only $f_1$, $f_2$, $f_6$, and $f_9$ were detected in all sectors, and $f_{11}$ in all but 
Sector 14. The five frequencies were stable within the standard deviation of $\sim$0.002 day$^{-1}$. The remaining frequencies 
except for some orbital harmonics ($f_{\rm orb}$, $2f_{\rm orb}$, $3f_{\rm orb}$) were not found in most sectors. 
On the other hand, if the high frequencies between 128 and 288 day$^{-1}$ result from WASP 1814+48 B with about 2 \% light 
contribution to the EB system, they are highly diluted by the light of WASP 1814+48 A. Then, these signals would have been 
too weak to detect in each sector with the frequency resolution of $\sim$0.055 day$^{-1}$. As the pre-He WD transits 
the larger and more massive primary star during the secondary eclipses, we analyzed the entire secondary-eclipse residuals 
(orbital phases 0.454$-$0.546). The primary-eclipse data were also analyzed for comparison. Figure 6 presents the amplitude 
spectra of both eclipse phases in the frequency region of 100$-$300 day$^{-1}$. Two high-frequency signals of 127.2627 and 
135.5987 day$^{-1}$ were detected only in the secondary eclipse phase. This implies that the main source of the high frequencies 
is the secondary companion WASP 1814+48 B.

\section{DISCUSSION AND CONCLUSIONS}

It is known from archival WASP photometry (Maxted et al. 2014) that WASP 1814+48 is a candidate EL CVn star. For the target star, 
we obtained the first spectroscopic observations using the BOES echelle spectrograph attached to the BOAO 1.8-m reflector. 
From the total 31 spectra, the RVs and atmospheric parameters of the cooler, more massive primary star were measured, and 
its surface temperature and rotation velocity were determined to be $T_{\rm eff,1}=7770 \pm 130$ K and $v_1$sin$i=47\pm6$ km s$^{-1}$, 
respectively. The spectroscopic measurements were solved with the high-precision photometric data from the TESS mission. 
The combined solution demonstrates that WASP 1814+48 is an EL CVn-type detached EB with masses of $M_1 = 1.659 \pm 0.048$ $M_\odot$ 
and $M_2 = 0.172 \pm 0.005$ $M_\odot$, radii of $R_1 = 1.945 \pm 0.027$ $R_\odot$ and $R_2 = 0.194 \pm 0.005$ $R_\odot$, and 
luminosities of $L_1 = 12.35 \pm 0.90$ $L_\odot$ and $L_2 = 0.69 \pm 0.07$ $L_\odot$. The component stars fill $f_1 = 49 \%$ 
and $f_2 = 36 \%$ of their inner Roche lobe, respectively. 

The surface gravity of the secondary companion can be calculated directly from the light and RV parameters without knowledge of 
its mass and radius (Southworth, Wheatley \& Sams 2007) and is represented as:
\begin{equation}
g_2 = {{G M_2} \over {R^2_2}} = {{2\pi} \over P} {{K_1(1-e^2)^{1/2}} \over {r^2_2\sin i}} 
\end{equation}
The observable quantities of $P$, $r_2$, $i$ and $K_1$ were taken from Table 4. This calculation results in $\log g_2$ = 5.097$\pm$0.025, 
which is well-matched with the 5.098$\pm$0.026 obtained from the secondary's mass and radius presented in this article. 

Following the solar values\footnote{($X_\odot$, $Y_\odot$, $Z_\odot$) = ($-$8.0, 0.0, 0.0) kpc and ($U_\odot$, $V_\odot$, 
$W_\odot$) = (9.58, 10.52, 7.01) km s$^{-1}$} and the procedure applied by Lee et al. (2020), the Galactic space motion of 
WASP 1814+48 was computed from our system velocity ($\gamma$) and the GAIA EDR3 measurements (position, parallax, and 
proper motion). The obtained velocity components were $U = -18.4 \pm 0.3$ km s$^{-1}$, $V = 252.8 \pm 0.6$ km s$^{-1}$, 
and $W = 9.1 \pm 0.3$ km s$^{-1}$, which correspond to a total space velocity of 253.6 $\pm 0.8$ km s$^{-1}$. Also, 
the Galactic orbit's eccentricity and angular momentum in the $z$ direction were obtained to be $e_{\rm G}$ = 
0.1813 $\pm$ 0.0003 and $J_{\rm z}$ = 1980 $\pm$ 17 kpc km s$^{-1}$, respectively. The position of WASP 1814+48 B in 
the $U-V$ and $J_{\rm z}-e_{\rm G}$ planes falls within the thin-disk population described by Pauli et al. (2006), indicating 
that our program target has the kinematics of thin-disk stars. 

Using the fundamental stellar parameters of WASP 1814+48, we studied the evolutionary history of the EB system in terms of 
the H-R and $\log T_{\rm eff}-\log g$ diagrams. The locations of the primary (A) and secondary (B) stars are presented as 
star symbols in Figure 7, while the oblique dash-dotted and dashed lines denote the instability strips of $\gamma$ Dor and 
$\delta$ Sct variables. WASP 1814+48 A resides inside the $\delta$ Sct region on the ZAMS, which implies that it is 
a $\delta$ Sct candidate. In both diagrams, the black dotted, dashed, and solid lines represent the evolutionary sequences 
of He-core WD stars with metallicities of $Z$ = 0.001, 0.01, and 0.02, respectively, for masses of $M$ = 0.182 $M_\odot$, 
0.176 $M_\odot$, and 0.171 $M_\odot$ (Istrate et al. 2016). We can see that WASP 1814+48 B with 0.172 $\pm$ 0.005 $M_\odot$ 
is in good agreement with the 0.176 $M_\odot$ WD model for $Z$ = 0.01. The result matches well with the thin-disk population 
classified by our Galactic kinematics. The lifetime ($t$) of the binary star in a constant luminosity phase was estimated to 
be about 1.2 $\times$ 10$^{9}$ yr using the $M_{\rm WD}-t$ relation (Chen et al. 2017). 

The whole light residuals from our binary model were analyzed using the software Period04 to find multiperiodic frequencies 
in our target star. We found two dominant oscillations at $f_{1}$ = 33.70479 day$^{-1}$ and $f_{2}$ = 32.80546 day$^{-1}$, 
corresponding to about 42.7 min and 43.9 min, respectively. As a consequence of the pre-whitening process, WASP 1814+48 
oscillated in a total of 52 frequencies that satisfied our criterion of S/N $\ga$ 4.0. Most signals in the low frequency region 
of $<$ 24 day$^{-1}$ may be sidelobes due to incomplete binary modeling and instrumental artifacts in the TESS observations, 
but not tidally excited modes and $\gamma$ Dor pulsations. The five frequencies between 32 and 36 day$^{-1}$ originated from 
WASP 1814+48 A located in the $\delta$ Sct instability domain, as shown in Figure 7. From the pulsation frequency-density relation 
of $Q_i$ = $f_i$$\sqrt{\rho / \rho_\odot}$, we determined their pulsation constants to be $Q_1$ = 0.014 days, $Q_2$ = 0.014 days, 
$Q_6$ = 0.015 days, $Q_9$ = 0.013 days, and $Q_{11}$ = 0.014 days in pressure ($p$) modes of $\delta$ Sct type with $Q < 0.04$ days 
(Breger 2000; Antoci et al. 2019). Moreover, the period ratios of $P_{\rm pul}/P_{\rm orb} = 0.0155 \sim 0.0171$ are within 
the threshold of 0.09 $\pm$ 0.02 for $\delta$ Sct EBs pulsating in $p$ modes (Zhang, Luo \& Fu 2013). 
On the other hand, the frequency signals between 128 and 288 day$^{-1}$ may be pulsation modes related to WASP 1814+48 B 
in the pre-He WD instability strip (C\'orsico et al. 2019). The periods and pulsation constants for the five high frequencies 
are in ranges of $5.0 < P_{\rm pul} < 11.2$ min and $0.017 \le Q \le 0.038$ days, respectively. These results make WASP 1814+48 
a very promising target for asteroseismology, consisting of a $\delta$ Sct-type primary and a pulsating pre-He WD companion.

\section*{Acknowledgments}
We would like to thank the BOAO staffs for assistance during our spectroscopy. This work includes data collected by the TESS 
mission, which were obtained from MAST. Funding for the TESS mission is provided by the NASA Explorer Program. This research 
was supported by the KASI grant 2022-1-830-04. K.H. was supported by the grants 2019R1A2C2085965 and 2020R1A4A2002885 from 
the National Research Foundation (NRF) of Korea.

\section*{DATA AVAILABILITY}
The data underlying this article will be shared on reasonable request to the first author.

\clearpage
\begin{figure}
\includegraphics{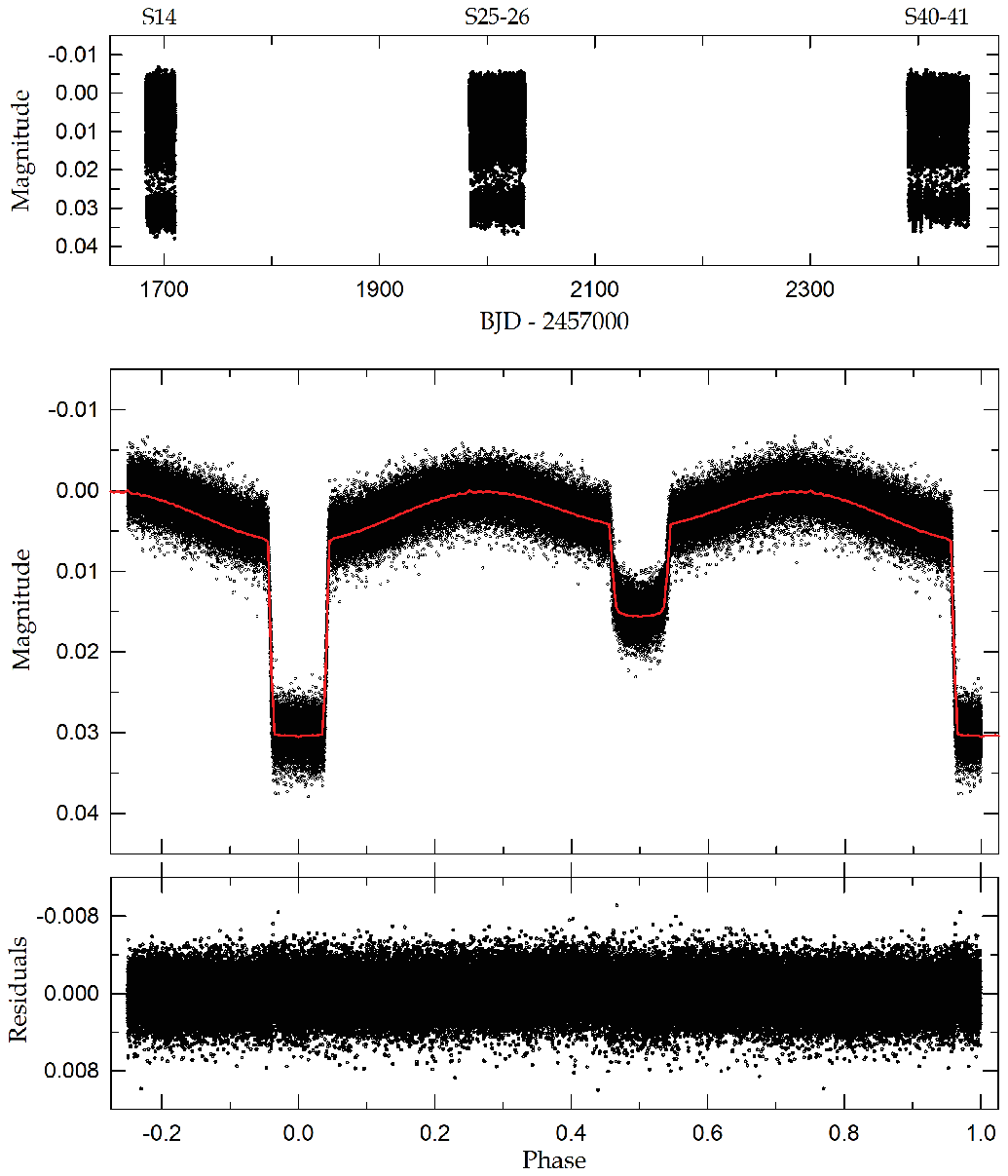}
\caption{TESS observations of WASP 1814+48 distributed in BJD (top panel) and orbital phase (middle panel). The circles are 
individual measures and the solid line represents the synthetic curve obtained with our binary model. The corresponding residuals 
are plotted in the bottom panel. } 
\label{Fig1}
\end{figure}

\begin{figure}
\includegraphics[]{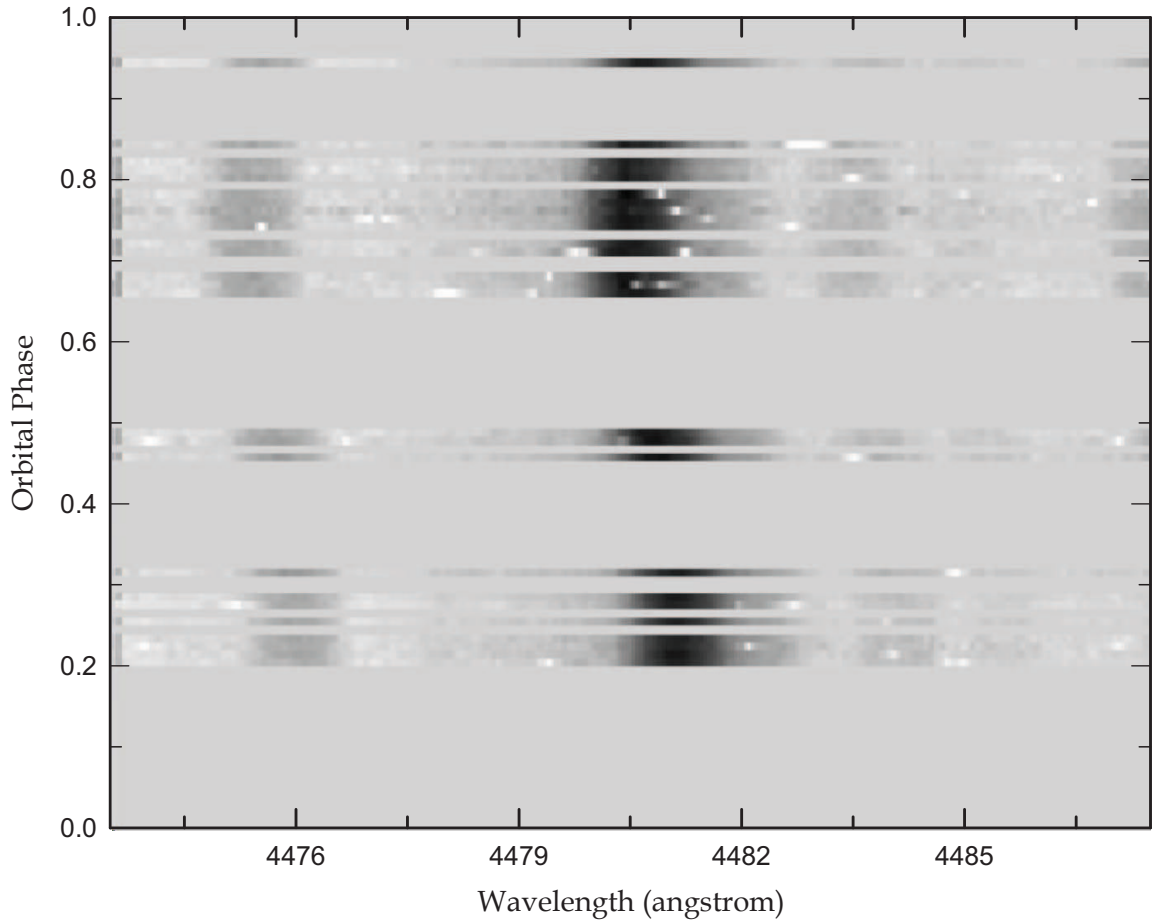}
\caption{Phase-folded trailed spectra of WASP 1814+48 in the Mg II $\lambda$4481 region. }
\label{Fig2}
\end{figure}

\begin{figure}
\includegraphics[]{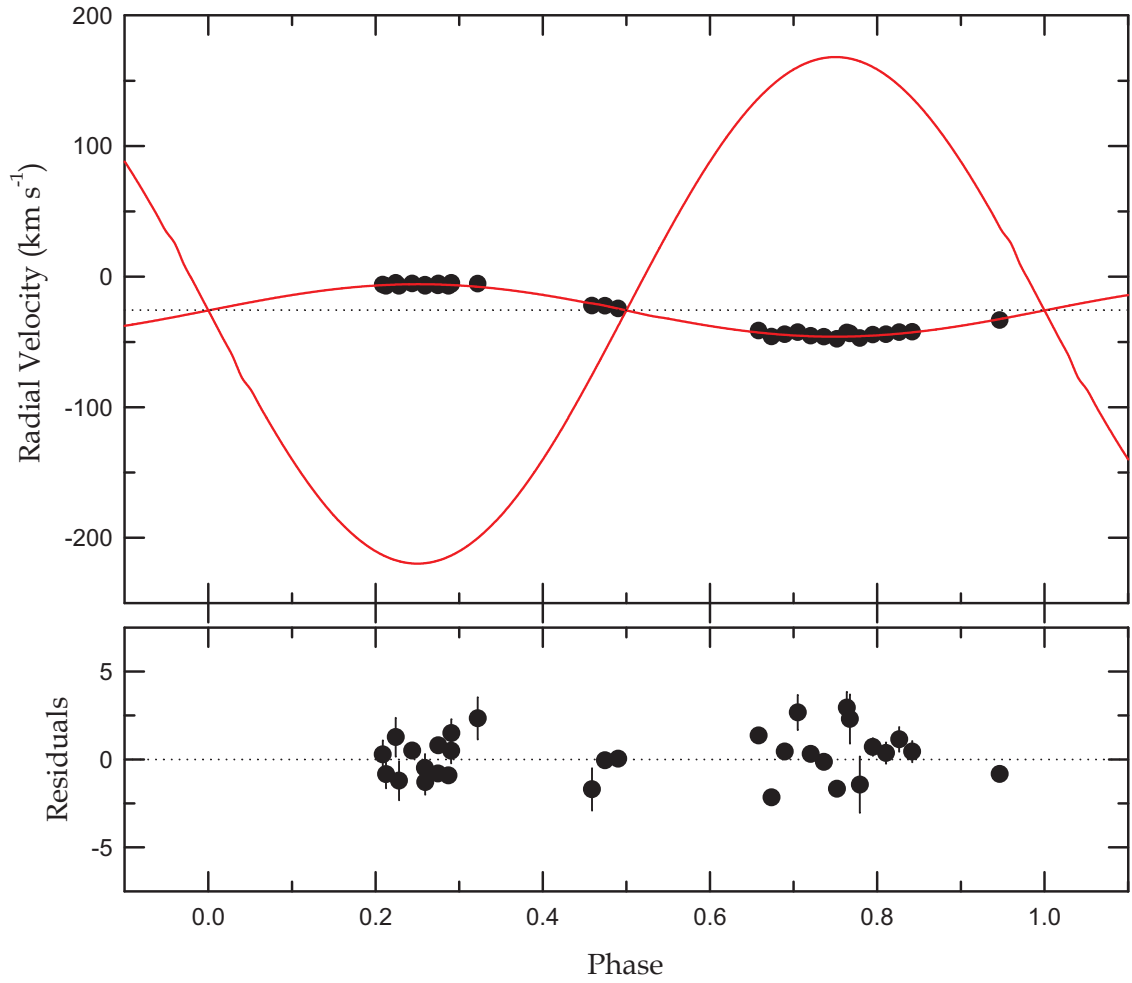}
\caption{Radial velocities of WASP 1814+48 with a fitted model. The solid curves represent the results from a consistent light 
and RV curve analysis and the dotted line denotes the system velocity of $-$25.65 km s$^{-1}$. The lower panel displays 
the residuals between measurements and models. }
\label{Fig3}
\end{figure}

\begin{figure}
\includegraphics{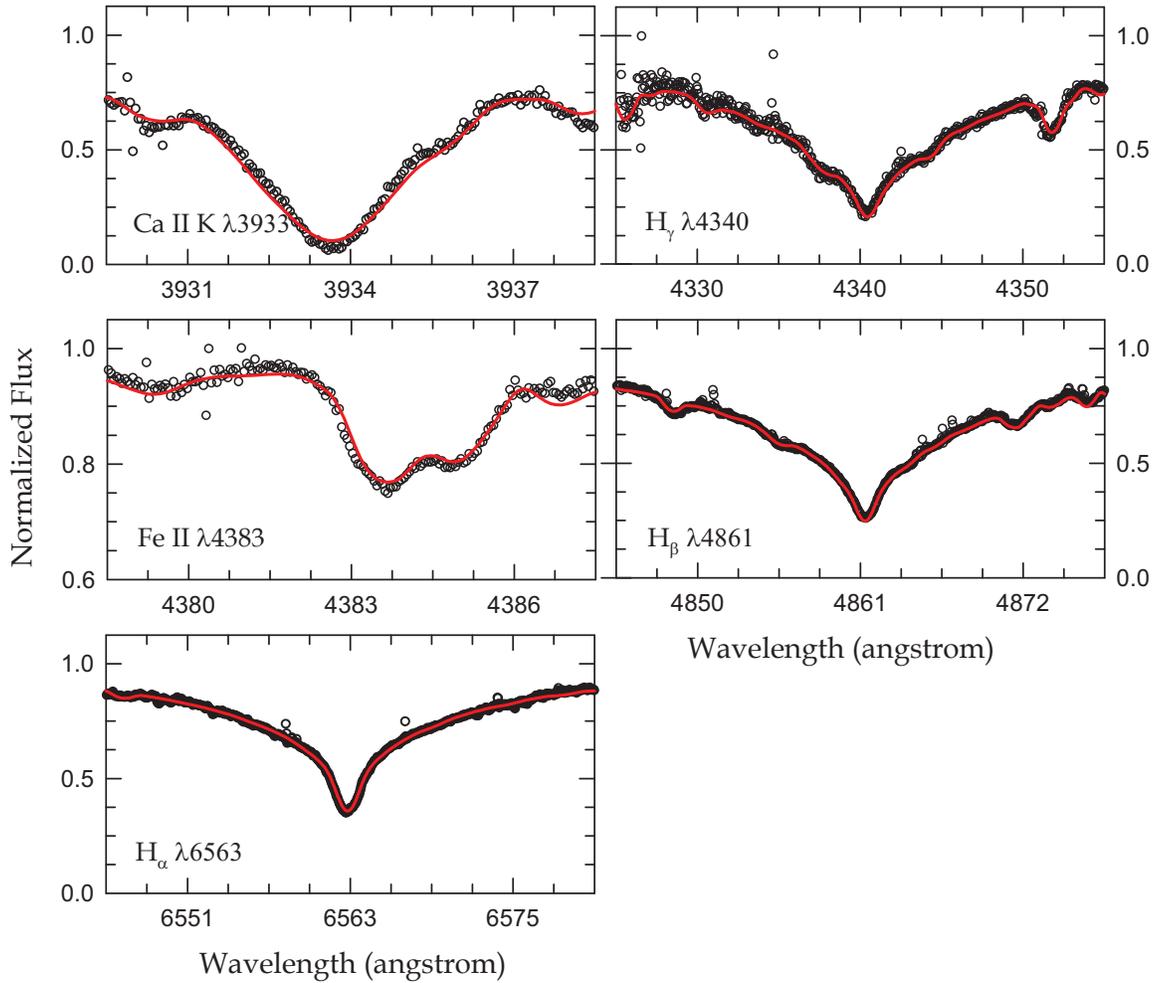}
\caption{Five spectral regions of WASP 1814+48 A. The open circles and the red lines represent the reconstructed spectrum 
obtained with the FDB\textsc{inary} code and the synthetic model spectrum with the best-fitting parameters of 
$T_{\rm eff,1}=7770 \pm 130$ K and $v_1$$\sin$$i=47\pm6$ km s$^{-1}$, respectively. The spectra were converted from vacuum 
to air wavelengths using the IAU standard conversion (Morton 1991). }
\label{Fig5}
\end{figure}

\begin{figure}
\includegraphics[]{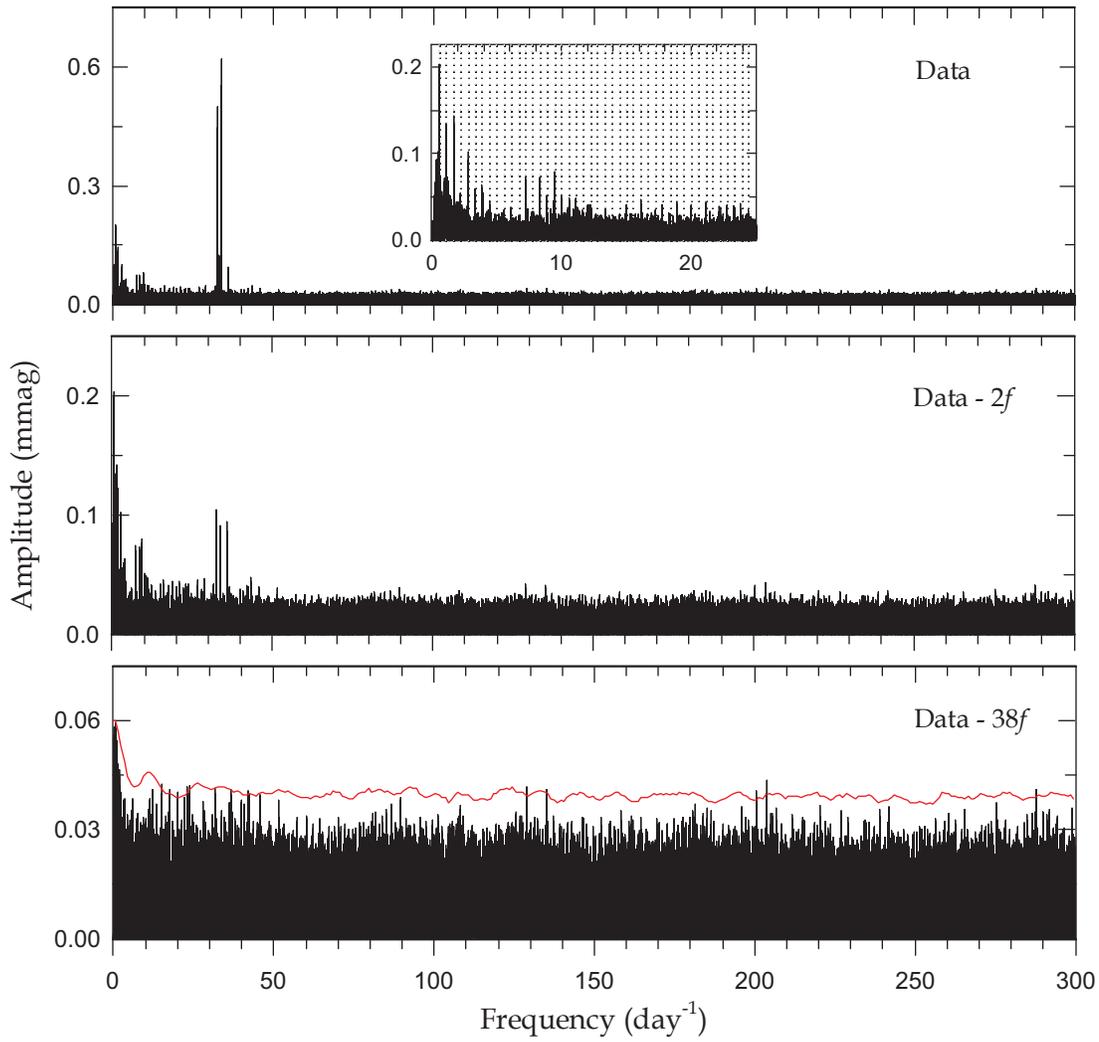}
\caption{Periodogram from the PERIOD04 program for the entire light curve residuals. The amplitude spectra before and after 
prewhitening the first two frequencies, and then 38 frequencies are shown in the top to bottom panels. The inset box in 
the top panel is the amplified spectrum of the low-frequency domain showing a series of orbital harmonics denoted by 
the vertical dotted lines. The red line in the bottom panel corresponds to four times the noise spectrum. }
\label{Fig5}
\end{figure}

\begin{figure}
\includegraphics[]{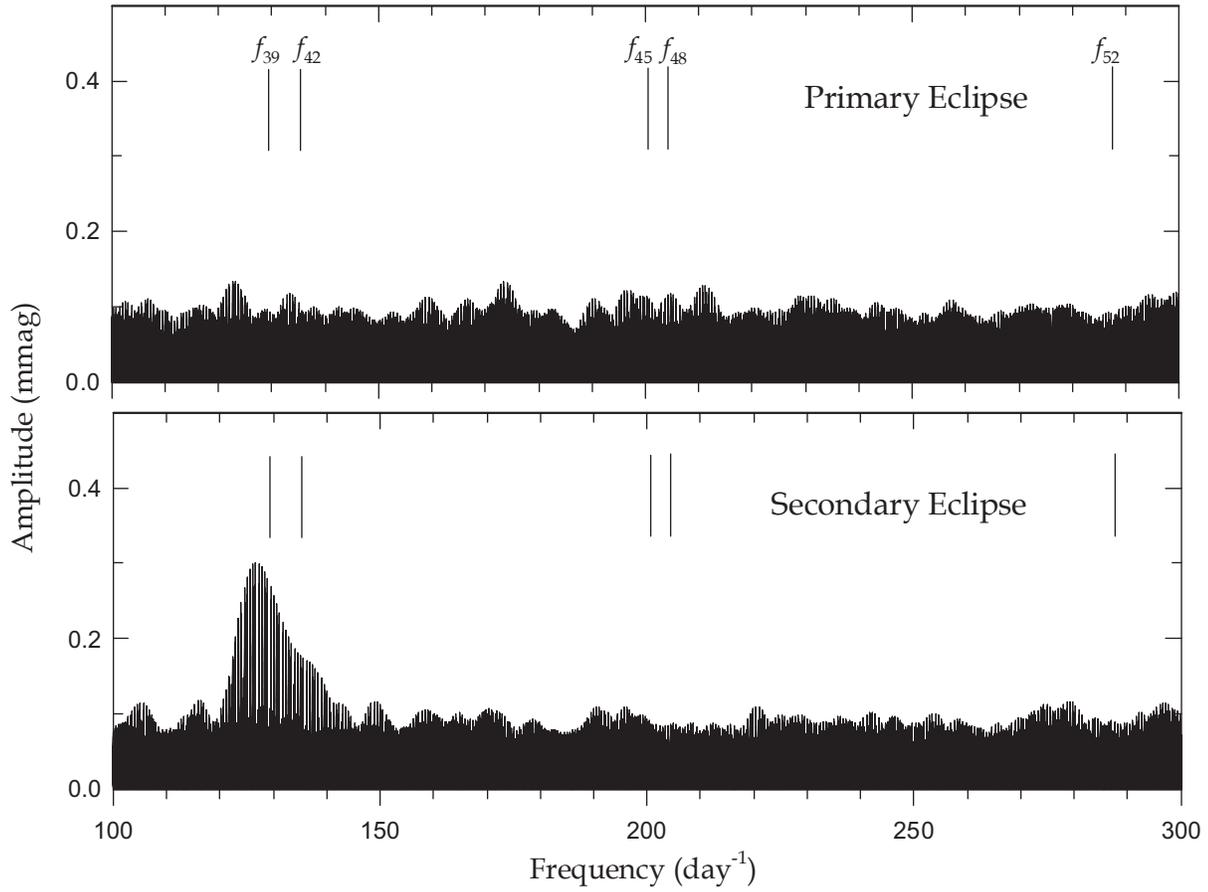}
\caption{Amplitude spectra for the light curve residuals during times of the primary (upper panel) and secondary (lower panel) 
eclipses. In both panels, the five vertical lines denote the pulsation frequencies of $f_{39}$, $f_{42}$, $f_{45}$, $f_{48}$, 
and $f_{52}$ found from the entire data listed in Table 6. }
\label{Fig6}
\end{figure}

\begin{figure}
\includegraphics{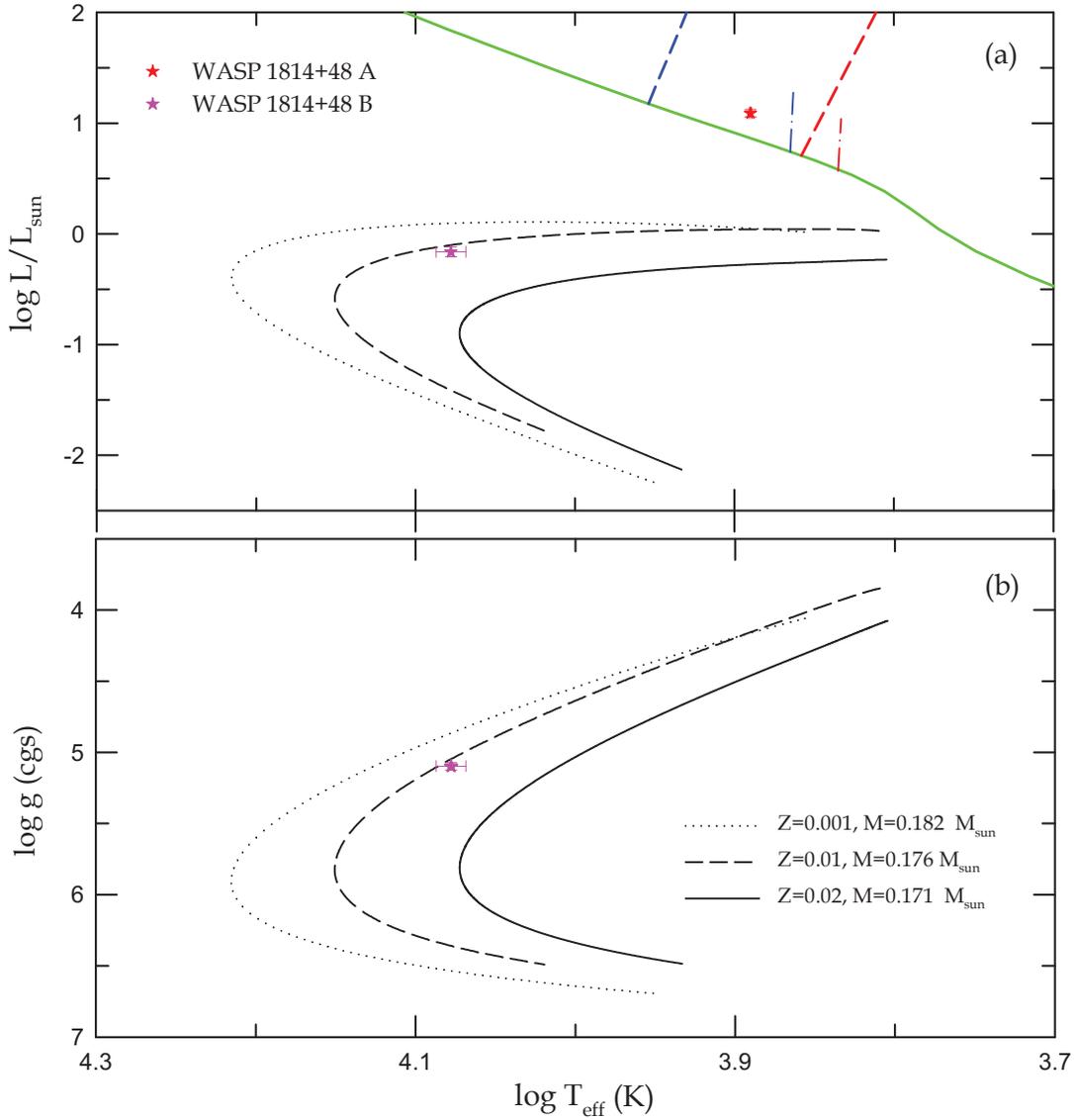}
\caption{(a) H-R and (b) $\log T_{\rm eff}-\log g$ diagrams for WASP 1814+48 (star symbols). In both panels, the black dotted, 
dashed, and solid lines are the evolutionary tracks of low-mass He WDs with metallicities $Z$ of 0.001, 0.01, and 0.02, 
respectively, for masses of 0.182 $M_\odot$, 0.176 $M_\odot$, and 0.171 $M_\odot$ (Istrate et al. 2016). In the upper panel (a), 
the green solid line denotes the ZAMS for solar metallicity, and the colored dashed and dash-dotted lines represent 
the instability strips of $\delta$ Sct and $\gamma$ Dor stars. } 
\label{Fig7}
\end{figure}

\clearpage
\begin{table}
\caption{TESS Eclipse Timings of WASP 1814+48. A sample is shown here: the full version is provided as supplementary material to the online article. }
\begin{tabular}{lcccc}
\hline
BJD              & Error           & Epoch           & $O-C$           & Min           \\                    
\hline                                         
2,458,683.90998  & $\pm$0.00057    & $-$181.5        & $-$0.00002      & II            \\
2,458,684.81013  & $\pm$0.00016    & $-$181.0        & $+$0.00041      & I             \\
2,458,685.70884  & $\pm$0.00038    & $-$180.5        & $-$0.00060      & II            \\
2,458,686.60921  & $\pm$0.00025    & $-$180.0        & $+$0.00006      & I             \\
2,458,687.50768  & $\pm$0.00028    & $-$179.5        & $-$0.00119      & II            \\
2,458,688.40861  & $\pm$0.00019    & $-$179.0        & $+$0.00003      & I             \\
2,458,689.30703  & $\pm$0.00033    & $-$178.5        & $-$0.00127      & II            \\
2,458,690.20880  & $\pm$0.00021    & $-$178.0        & $+$0.00079      & I             \\
2,458,691.10946  & $\pm$0.00025    & $-$177.5        & $+$0.00173      & II            \\
2,458,692.00737  & $\pm$0.00029    & $-$177.0        & $-$0.00008      & I             \\
\hline
\end{tabular}
\end{table}

\begin{table}
\caption{Radial Velocities of WASP 1814+48 A. }
\begin{tabular}{lrcr}
\hline
BJD                    & $V_{1}$                 & $\sigma_1$              & $O-C$                    \\ 
(2,450,000+)           & (km s$^{-1}$)           & (km s$^{-1}$)           & (km s$^{-1}$)            \\                    
\hline                                         
7,128.2085             & $ -33.3 $               & 0.3                     & $-0.8$                   \\ 
8,963.1064             & $ -41.4 $               & 0.3                     & $ 1.4$                   \\ 
8,963.1344             & $ -45.9 $               & 0.4                     & $-2.2$                   \\ 
8,963.1625             & $ -44.1 $               & 0.4                     & $ 0.4$                   \\ 
8,963.1906             & $ -42.5 $               & 1.0                     & $ 2.7$                   \\ 
8,963.2186             & $ -45.3 $               & 0.3                     & $ 0.3$                   \\ 
8,963.2467             & $ -46.0 $               & 0.3                     & $-0.1$                   \\ 
8,963.2748             & $ -47.6 $               & 0.3                     & $-1.7$                   \\ 
8,963.3029             & $ -43.5 $               & 1.4                     & $ 2.3$                   \\ 
8,964.0968             & $  -6.2 $               & 0.8                     & $ 0.3$                   \\ 
8,964.1249             & $  -4.8 $               & 1.1                     & $ 1.3$                   \\ 
8,964.1876             & $  -6.3 $               & 0.8                     & $-0.5$                   \\ 
8,964.2157             & $  -6.8 $               & 0.4                     & $-0.8$                   \\ 
8,964.2438             & $  -5.9 $               & 0.7                     & $ 0.5$                   \\ 
8,964.3011             & $  -5.4 $               & 1.2                     & $ 2.3$                   \\ 
8,968.1461             & $ -22.2 $               & 1.2                     & $-1.7$                   \\ 
8,968.1742             & $ -22.3 $               & 0.3                     & $-0.1$                   \\ 
8,968.2023             & $ -24.4 $               & 0.4                     & $ 0.0$                   \\ 
9,000.2267             & $  -7.2 $               & 0.4                     & $-0.9$                   \\ 
9,304.1956             & $  -7.2 $               & 0.8                     & $-0.8$                   \\ 
9,304.2237             & $  -7.2 $               & 1.1                     & $-1.2$                   \\ 
9,304.2518             & $  -5.3 $               & 0.3                     & $ 0.5$                   \\ 
9,304.2799             & $  -7.1 $               & 0.7                     & $-1.3$                   \\ 
9,304.3081             & $  -5.2 $               & 0.3                     & $ 0.8$                   \\ 
9,304.3362             & $  -4.9 $               & 0.8                     & $ 1.5$                   \\ 
9,305.1875             & $ -42.9 $               & 0.9                     & $ 2.9$                   \\ 
9,305.2156             & $ -47.0 $               & 1.6                     & $-1.4$                   \\ 
9,305.2437             & $ -44.4 $               & 0.5                     & $ 0.7$                   \\ 
9,305.2719             & $ -44.1 $               & 0.6                     & $ 0.4$                   \\ 
9,305.3000             & $ -42.5 $               & 0.7                     & $ 1.1$                   \\ 
9,305.3281             & $ -42.2 $               & 0.6                     & $ 0.4$                   \\ 
\hline
\end{tabular}
\end{table}

\begin{table}
\caption{Orbital Elements of WASP 1814+48. }
\begin{tabular}{lc}
\hline
Parameter                     & Value                         \\
\hline  
$T_0$ (HJD)                   & 2,459,010.5067$\pm$0.0054     \\
$P$ (day)$\rm ^a$             & 1.79943078                    \\
$\gamma$ (km s$^{-1}$)        & $-25.85\pm$0.20               \\
$K_1$ (km s$^{-1}$)           & 20.17$\pm$0.23                \\ 
$a_1$$\sin$$i$ ($R_\odot$)    & 0.717$\pm$0.010               \\
$f$(M) ($M_\odot$)            & 0.00153$\pm$0.00010           \\
rms (km s$^{-1}$)             & 1.35                          \\
\hline
\multicolumn{2}{l}{$^a$ Fixed.} \\
\end{tabular}
\end{table}

\begin{table}
\caption{Light and Velocity Parameters of WASP 1814+48. }
\begin{tabular}{lcc}
\hline
Parameter                         & Primary            & Secondary                    \\ 
\hline                                                                                      
$T_0$ (HJD)                       & \multicolumn{2}{c}{2,459,010.50670$\pm$0.00023}   \\
$P$ (day)                         & \multicolumn{2}{c}{1.7994280$\pm$0.0000014}       \\
$a$ (R$_\odot$)                   & \multicolumn{2}{c}{7.611$\pm$0.095}               \\
$\gamma$ (km s$^{-1}$)            & \multicolumn{2}{c}{$-25.65\pm$0.22}               \\
$K_1$ (km s$^{-1}$)               & \multicolumn{2}{c}{20.13$\pm$0.23}                \\
$K_2$ (km s$^{-1}$)               & \multicolumn{2}{c}{193.9$\pm$2.7}                 \\
$q$                               & \multicolumn{2}{c}{0.1038$\pm$0.0019}             \\
$i$ (deg)                         & \multicolumn{2}{c}{89.27$\pm$0.13}                \\
$T$ (K)                           & 7770$\pm$130       & 11,964$\pm$256               \\
$\Omega$                          & 4.041$\pm$0.022    & 5.522$\pm$0.041              \\
$\Omega_{\rm in}$$\rm ^a$         & \multicolumn{2}{c}{1.971}                         \\
$f$ (\%)$\rm ^b$                  & 48.8               & 35.7                         \\
$X$, $Y$                          & 0.664, 0.226       & 0.737, 0.075                 \\
$x$, $y$                          & 0.506, 0.213       & 0.371, 0.186                 \\
$l/(l_1+l_2)$                     & 0.9784$\pm$0.0005  & 0.0216                       \\
$r$ (pole)                        & 0.2538$\pm$0.0015  & 0.0255$\pm$0.0006            \\
$r$ (point)                       & 0.2570$\pm$0.0016  & 0.0255$\pm$0.0006            \\
$r$ (side)                        & 0.2561$\pm$0.0016  & 0.0255$\pm$0.0006            \\
$r$ (back)                        & 0.2567$\pm$0.0016  & 0.0255$\pm$0.0006            \\
$r$ (volume)$\rm ^c$              & 0.2556$\pm$0.0016  & 0.0255$\pm$0.0006            \\ 
\hline
\multicolumn{3}{l}{$^a$ Potential for the inner critical Roche surface.} \\
\multicolumn{3}{l}{$^b$ Fill-out factor $f = \Omega_{\rm in} / \Omega$ $\times$ 100.} \\
\multicolumn{3}{l}{$^c$ Mean volume radius.} 
\end{tabular}
\end{table}

\begin{table}
\caption{Absolute Parameters of WASP 1814+48. }
\begin{tabular}{lcc}
\hline
Parameter                     & Primary             & Secondary                   \\                                                                                         
\hline 
$M$ ($M_\odot$)               & 1.659$\pm$0.048     & 0.172$\pm$0.005             \\
$R$ ($R_\odot$)               & 1.945$\pm$0.027     & 0.194$\pm$0.005             \\
$\log$ $g$ (cgs)              & 4.080$\pm$0.018     & 5.098$\pm$0.026             \\
$\rho$ ($\rho_\odot$)         & 0.226$\pm$0.012     & 23.6$\pm$2.0                \\
$v_{\rm sync}$ (km s$^{-1}$)  & 54.70$\pm$0.77      & 5.46$\pm$0.15               \\
$v$$\sin$$i$ (km s$^{-1}$)    & 47$\pm$6            & \,                          \\
$T_{\rm eff}$ (K)             & 7770$\pm$130        & 11,964$\pm$256              \\
$L$ ($L_\odot$)               & 12.35$\pm$0.90      & 0.69$\pm$0.07               \\
$M_{\rm bol}$ (mag)           & 2.00$\pm$0.08       & 5.13$\pm$0.11               \\
BC (mag)                      & 0.03$\pm$0.01       & $-$0.68$\pm$0.05            \\
$M_{\rm V}$ (mag)             & 1.97$\pm$0.08       & 5.81$\pm$0.12               \\
Distance (pc)                 & \multicolumn{2}{c}{536$\pm$20}                    \\
\hline
\end{tabular}
\end{table}

\begin{table}
\caption{Results of the multiple frequency analysis for WASP 1814+48$\rm ^a$. }
\begin{tabular}{lrccrc}
\hline
             & Frequency              & Amplitude           & Phase           & S/N$\rm ^b$    & Remark                     \\
             & (day$^{-1}$)           & (mmag)              & (rad)           &                &                            \\
\hline
$f_{1}$      &  33.70479$\pm$0.00001  & 0.618$\pm$0.018     & 3.23$\pm$0.08   & 59.44          &                            \\
$f_{2}$      &  32.80546$\pm$0.00001  & 0.504$\pm$0.018     & 4.55$\pm$0.10   & 48.02          &                            \\
$f_{3}$      &   0.55558$\pm$0.00003  & 0.221$\pm$0.026     & 2.44$\pm$0.34   & 14.62          & $1f_{\rm orb}$             \\
$f_{4}$      &   1.66715$\pm$0.00005  & 0.136$\pm$0.024     & 5.43$\pm$0.52   &  9.61          & $3f_{\rm orb}$             \\
$f_{5}$      &   1.10969$\pm$0.00004  & 0.156$\pm$0.025     & 4.48$\pm$0.47   & 10.58          & $2f_{\rm orb}$             \\
$f_{6}$      &  32.59328$\pm$0.00005  & 0.104$\pm$0.018     & 4.36$\pm$0.51   &  9.92          &                            \\
$f_{7}$      &   2.77858$\pm$0.00006  & 0.104$\pm$0.022     & 0.35$\pm$0.63   &  7.95          & $5f_{\rm orb}$             \\
$f_{8}$      &   0.45251$\pm$0.00008  & 0.090$\pm$0.026     & 4.73$\pm$0.85   &  5.93          &                            \\                 
$f_{9}$      &  35.86918$\pm$0.00005  & 0.096$\pm$0.018     & 1.91$\pm$0.54   &  9.21          &                            \\
$f_{10}$     &   0.49203$\pm$0.00008  & 0.088$\pm$0.026     & 0.53$\pm$0.86   &  5.82          &                            \\                 
$f_{11}$     &  33.91699$\pm$0.00005  & 0.091$\pm$0.018     & 3.67$\pm$0.58   &  8.59          &                            \\
$f_{12}$     &   0.39523$\pm$0.00008  & 0.087$\pm$0.026     & 5.55$\pm$0.88   &  5.71          &                            \\                 
$f_{13}$     &   0.50368$\pm$0.00009  & 0.078$\pm$0.026     & 1.19$\pm$0.98   &  5.11          &                            \\                 
$f_{14}$     &   9.44726$\pm$0.00006  & 0.079$\pm$0.019     & 2.19$\pm$0.70   &  7.15          & $17f_{\rm orb}$            \\ 
$f_{15}$     &   7.22447$\pm$0.00006  & 0.074$\pm$0.018     & 6.25$\pm$0.71   &  7.03          & $13f_{\rm orb}$            \\ 
$f_{16}$     &   8.33590$\pm$0.00006  & 0.074$\pm$0.018     & 5.07$\pm$0.72   &  6.98          & $15f_{\rm orb}$            \\ 
$f_{17}$     &   0.47818$\pm$0.00008  & 0.083$\pm$0.026     & 4.40$\pm$0.91   &  5.49          &                            \\                 
$f_{18}$     &   0.93935$\pm$0.00011  & 0.061$\pm$0.025     & 3.36$\pm$1.19   &  4.21          &                            \\                 
$f_{19}$     &   1.11036$\pm$0.00005  & 0.129$\pm$0.025     & 4.06$\pm$0.57   &  8.72          & $2f_{\rm orb}$             \\ 
$f_{20}$     &   0.23589$\pm$0.00010  & 0.069$\pm$0.026     & 3.84$\pm$1.12   &  4.48          &                            \\                 
$f_{21}$     &   3.89005$\pm$0.00008  & 0.063$\pm$0.020     & 4.52$\pm$0.94   &  5.32          & $7f_{\rm orb}$             \\ 
$f_{22}$     &   0.98233$\pm$0.00011  & 0.064$\pm$0.026     & 5.72$\pm$1.18   &  4.26          & $2f_{10}$                  \\           
$f_{23}$     &   3.33800$\pm$0.00009  & 0.062$\pm$0.021     & 1.47$\pm$1.01   &  4.96          & $6f_{\rm orb}$(?)          \\ 
$f_{24}$     &   1.23523$\pm$0.00011  & 0.063$\pm$0.026     & 4.96$\pm$1.21   &  4.13          & $2f_5-f_{22}$              \\               
$f_{25}$     &   0.29531$\pm$0.00011  & 0.061$\pm$0.026     & 3.10$\pm$1.26   &  3.99          &                            \\
$f_{26}$     &   0.34236$\pm$0.00010  & 0.067$\pm$0.026     & 1.22$\pm$1.14   &  4.41          &                            \\
$f_{27}$     &   2.22238$\pm$0.00011  & 0.056$\pm$0.023     & 4.33$\pm$1.23   &  4.08          & $4f_{\rm orb}$             \\
$f_{28}$     &  10.00365$\pm$0.00010  & 0.051$\pm$0.019     & 2.32$\pm$1.12   &  4.47          & $18f_{\rm orb}$            \\
$f_{29}$     &   8.89224$\pm$0.00010  & 0.050$\pm$0.019     & 2.22$\pm$1.09   &  4.58          & $16f_{\rm orb}$            \\
$f_{30}$     &  10.55889$\pm$0.00010  & 0.049$\pm$0.020     & 2.51$\pm$1.17   &  4.30          & $19f_{\rm orb}$            \\
$f_{31}$     &  11.11509$\pm$0.00011  & 0.048$\pm$0.020     & 2.89$\pm$1.20   &  4.17          & $20f_{\rm orb}$            \\
$f_{32}$     &  43.34495$\pm$0.00009  & 0.047$\pm$0.017     & 0.48$\pm$1.04   &  4.82          &                            \\
$f_{33}$     &  28.89502$\pm$0.00010  & 0.047$\pm$0.018     & 4.15$\pm$1.12   &  4.49          & $f_{32}-26f_{\rm orb}$     \\
$f_{34}$     &  16.11616$\pm$0.00010  & 0.046$\pm$0.017     & 5.67$\pm$1.09   &  4.60          & $29f_{\rm orb}$            \\
$f_{35}$     &  26.67267$\pm$0.00011  & 0.045$\pm$0.018     & 4.20$\pm$1.19   &  4.20          & $f_{32}-30f_{\rm orb}$     \\
$f_{36}$     &   4.45508$\pm$0.00011  & 0.045$\pm$0.019     & 2.07$\pm$1.27   &  3.96          &                            \\
$f_{37}$     &  18.89540$\pm$0.00010  & 0.045$\pm$0.017     & 2.30$\pm$1.10   &  4.55          & $34f_{\rm orb}$            \\
$f_{38}$     &  21.12724$\pm$0.00010  & 0.044$\pm$0.017     & 6.11$\pm$1.10   &  4.56          &                            \\
$f_{39}$     & 203.85745$\pm$0.00010  & 0.044$\pm$0.017     & 1.85$\pm$1.13   &  4.43          &                            \\
$f_{40}$     &  15.00015$\pm$0.00011  & 0.042$\pm$0.018     & 5.37$\pm$1.23   &  4.07          & $27f_{\rm orb}$(?)         \\
$f_{41}$     &  23.84905$\pm$0.00011  & 0.042$\pm$0.018     & 2.34$\pm$1.23   &  4.09          &                            \\
$f_{42}$     & 128.93254$\pm$0.00011  & 0.042$\pm$0.017     & 1.40$\pm$1.19   &  4.22          &                            \\
$f_{43}$     &  31.57909$\pm$0.00011  & 0.042$\pm$0.018     & 3.19$\pm$1.25   &  4.00          &                            \\
$f_{44}$     &  22.78744$\pm$0.00011  & 0.042$\pm$0.017     & 3.01$\pm$1.21   &  4.16          & $41f_{\rm orb}$(?)         \\
$f_{45}$     & 135.25945$\pm$0.00011  & 0.041$\pm$0.017     & 3.53$\pm$1.22   &  4.11          &                            \\
$f_{46}$     &  36.54663$\pm$0.00011  & 0.041$\pm$0.018     & 1.60$\pm$1.25   &  4.02          &                            \\
$f_{47}$     &  23.33990$\pm$0.00011  & 0.041$\pm$0.017     & 1.90$\pm$1.23   &  4.07          & $42f_{\rm orb}$            \\
$f_{48}$     & 287.85952$\pm$0.00011  & 0.041$\pm$0.017     & 3.45$\pm$1.18   &  4.24          &                            \\
$f_{49}$     &  17.78415$\pm$0.00011  & 0.041$\pm$0.017     & 6.20$\pm$1.22   &  4.12          & $32f_{\rm orb}$            \\
$f_{50}$     &  42.23532$\pm$0.00011  & 0.041$\pm$0.017     & 0.07$\pm$1.22   &  4.12          & $f_{32}-2f_{\rm orb}$      \\
$f_{51}$     & 200.42459$\pm$0.00010  & 0.040$\pm$0.016     & 2.74$\pm$1.18   &  4.26          &                            \\
$f_{52}$     &  20.00649$\pm$0.00011  & 0.040$\pm$0.016     & 3.73$\pm$1.20   &  4.18          & $36f_{\rm orb}$            \\
\hline 
\multicolumn{6}{l}{$^a$ Frequencies are listed in order of detection.} \\
\multicolumn{6}{l}{$^b$ Calculated in a range of 5 day$^{-1}$ around each frequency.} 
\end{tabular}
\end{table}

\bsp
\label{lastpage}
\end{document}